\documentclass[12pt]{iopart}
\usepackage{graphicx,bm}
\usepackage{iopams}  
\usepackage{bm}

\newcommand{\be}{\begin{eqnarray}}
\newcommand{\ee}{\end{eqnarray}}
\newcommand{\no}{\nonumber}
\newcommand{\lb}{\left(}
\newcommand{\rb}{\right)}
\newcommand{\lbb}{\left\{}
\newcommand{\rbb}{\right\}}
\newcommand{\Part}[3]{\frac{\partial^{#3}#1}{\partial #2^{#3}}}

\begin{document}

\title[Zeros of spherical spin glasses and chaos]
{Partition-function zeros of spherical spin glasses and their relevance to chaos}

\author{Tomoyuki Obuchi$^1$, Kazutaka Takahashi$^2$}

\address{
$^1$ Department of Earth and Space Science, Faculty of Science, \\
 Osaka University, Osaka 560-0043, Japan

$^2$ Department of Physics, 
 Tokyo Institute of Technology, Tokyo 152-8551, Japan
}
\begin{abstract}
We investigate partition-function zeros of 
the many-body interacting spherical spin glass, the so-called $p$-spin spherical model, 
with respect to the complex temperature in the thermodynamic limit. 
We use the replica method and 
extend the procedure of the replica symmetry breaking ansatz
to be applicable in the complex-parameter case. 
We derive the phase diagrams in the complex-temperature plane 
and calculate the density of zeros in each phase. 
Near the imaginary axis away from the origin, 
there is a replica symmetric phase having a large density. 
On the other hand, we observe no density in the spin-glass 
phases, irrespective of the replica symmetry breaking. 
We speculate that this suggests the absence of the temperature chaos. 
To confirm this, we investigate the multiple many-body interacting case which is known to exhibit the chaos effect. 
The result shows that the density of zeros actually takes 
finite values in the spin-glass phase, even on the real axis. 
These observations indicate that the density of zeros is more closely 
connected to the chaos effect than the replica symmetry breaking. 
\end{abstract}

\maketitle

\section{Introduction}
\label{sec:introduction}
Phase transitions and critical phenomena have been 
a central problem in statistical physics for decades. 
After many pioneering works, it was revealed that phase transitions 
can be identified as singularities of the free energy, 
and several approaches to capture them 
have also been investigated for a long 
time~\cite{Onsager:44,Kaufmann:49,Yang:52-1,Yang:52-2,Fisher:64}. 
The theory of partition-function zeros invented 
by Yang and Lee~\cite{Yang:52-1,Yang:52-2} is one of such approaches 
and offers a novel, and simple, picture of phase transitions. 
They proved that the free energy is analytic in a region where 
there are no zeros of the partition function, 
and hence there is no phase transition in that region. 
Besides, using Ising ferromagnets, they demonstrated that the phase 
transitions of the models become clearly visible by the zeros. 
Their work was followed by many other researchers and 
was applied to various situations~\cite{Fisher:64,Bena:05}

Spin glass (SG) is known to show nontrivial phase transitions and 
critical phenomena and has been studied for a long time~\cite{SPIN,STAT}. 
According to the standard description of SGs, 
a SG system acquires a multi-valley structure in the free energy 
landscape at low temperatures.
Some peculiar properties associated with SG transitions, 
such as strong hysteresis and rejuvenation-memory effect, 
are explained on the basis of this rugged landscape. 
In the mean-field level, each valley of the free energy is 
separated by infinitely-high free-energy barriers and 
is called a pure state. 
Each pure state corresponds to a thermodynamic phase. 
This provides a speculation that a sequence of phase transitions 
in a sense can occur in SG phases where the dominant part of pure states 
can vary as we change external parameters such as temperature. 
Unfortunately, it is difficult to directly examine this kind of 
transitions in the mean-field solution given 
by Parisi~\cite{Parisi:80-1,Parisi:80-2}. 
This motivates us to use another approach, the zeros theory by Yang and Lee. 

Zeros of SGs have also been investigated for a fairly long 
time~\cite{Ozeki:88,Faria:91,Bhanot:93,Damgaard:95,
Matsuda:08,Derrida:91,Moukarzel:91,Moukarzel:92-1,Moukarzel:92-2}, 
but until very recently the reasonable solution in the thermodynamic limit 
was not obtained even for mean-field models~\cite{Matsuda:10,Takahashi:11}, 
except for a simple model named the random energy model 
(REM)~\cite{Derrida:91,Moukarzel:92-1}. 
Observing these few solutions in the thermodynamic 
limit~\cite{Derrida:91,Matsuda:10,Takahashi:11}, 
we find a tendency that the distributions of zeros are closely related 
to the step number of the replica symmetry breaking (RSB) 
in the Parisi scheme. 
For Bethe SGs exhibiting the full-step RSB (FRSB), 
the zeros tend to densely distribute around the real axes of the temperature 
and uniform field below the critical points~\cite{Matsuda:10}. 
This suggests that a certain type of phase transitions occur everywhere 
in the FRSB phase, which supports the above speculation. 
This type of transitions can be possibly interpreted as 
the temperature/field chaos meaning that the spin configuration 
drastically changes as the temperature/field slightly 
varies~\cite{Parisi:83,Binder:86,Kondor:89,Kondor:93,Franz:95,Rizzo:01,Rizzo:02,Rizzo:03,Rizzo:06}. 
On the other hand, for a family of REMs exhibiting 
the one-step RSB (1RSB), no zeros in the complex temperature plane, or very few zeros in the complex field plane, 
exist in the internal region of 
the SG phases~\cite{Derrida:91,Moukarzel:92-1,Takahashi:11}. 
These contrasting results possibly reflect the difference 
between the 1RSB and FRSB, or may be just due to 
the peculiarity of the REMs being over-simplified models. 
To make this point clear, we need to analyze distributions of zeros 
in more realistic SG models, which is the main purpose of this paper.

Although the validity of the RSB picture is questioned 
in finite-dimensional SGs~\cite{Fisher:86,Bray:87,Fisher:88,Reger:90,Moore:98,Kawashima:00,Marinari:00-1,Marinari:00-2,Drossel:00-1,Drossel:00-2,Hartmann:02,Jorg:08,Banos:10}, the temperature chaos is considered to exist in those systems~\cite{Bray:87,McKay:82,Banavar:87,Ney-Nifle:98,Aspelmeier:02,Sasaki:05,Katzgraber:08}. 
Hence, it will be helpful to reveal the relation between the chaos effect and 
the density of zeros (DOZ).
Up to the present, we have no fully-reliable result about the zeros of 
finite-dimensional SGs in the thermodynamic limit, 
though they are intensively studied~\cite{Ozeki:88,Bhanot:93,Damgaard:95,Matsuda:08}. 
We expect that some clear knowledge about the DOZ of SGs, 
even in the mean-field level, can be a help to improve this situation. 

In this paper, we investigate the distribution of zeros of 
the many-body interacting spherical SGs~\cite{Kosterlitz:76,Crisanti:92}. These are more natural than the REMs in that the phase-space decomposition into many pure states occurs as the temperature changes. 
Besides, it shows both the replica symmetric (RS) SG phase and 
the 1RSB-SG phase depending on a parameter $p$ being the number 
of interacting spins. 
Moreover, it is easy to control the temperature chaos of this model. 
In the case of the single $p$-body interaction, 
there is no temperature chaos. 
However, in the multiple $(p+r)$-body interacting case 
it is known that the temperature chaos occurs~\cite{Rizzo:06}. 
These properties are quite useful to investigate the relations 
among the DOZ, the RSB and the chaos effect. 

To derive the DOZ, we use the formulation invented 
in~\cite{Takahashi:11}. 
In that formulation, we employ the replica method and 
generalize the Parisi scheme to be applicable 
in the complex-parameter cases. 
Three different types of overlaps between replicas are introduced. 
The physical interpretation of the overlaps is also one of 
the results in this paper. 
Although we concentrate only on the zeros in the complex temperature plane, 
our formulation can be applied to 
the complex field or other physical parameters.

This paper is organized as follows. 
In the next section, we start from a brief introduction of partition-function zeros. The replica-based formulation to assess DOZs is 
also explained in this section. 
In section~\ref{sec:Replica}, we introduce the spherical SG model and 
derive the saddle-point equations to calculate the DOZ in the replica formulation. 
The RS and 1RSB solutions are investigated and 
the physical significances are discussed. 
In section~\ref{sec:diagrams}, we present the phase diagrams 
in the complex temperature plane and the values of the DOZ in each phase. 
Last section is devoted to conclusion.

\section{Formulation}
\label{sec:formulations}

\subsection{Partition-function zeros}
\label{sec:zeros}

Since a partition function of a finite size system is generally 
analytic with respect to a physical parameter $y$, 
we can reasonably assume that 
the partition function of the size $N$ can be factorized 
into a product form 
\be
Z(y)=e^{NC}\prod_{j}(y-y^{(j)}),
\ee
where $\{y^{(j)}\}$ are zeros of the partition function and 
generally complex $y^{(j)}=y_1^{(j)}+iy_2^{(j)}$, 
where $i$ is the imaginary unit. 
We assume that $C$ is an analytic and irrelevant factor.
Hence, the free energy density of the system
$f= -(N\beta)^{-1}\ln Z$ is written by
\be
-\beta f(y)=C+\sum_{j}\frac{1}{N}\ln(y-y^{(j)})
=C+\int dz_1 dz_2 \rho(z_1,z_2)\ln(y-z),
\ee
where $z=z_1+iz_2$ and we define the DOZ $\rho(z_1,z_2)$ as
\be
\rho(z_1,z_2)=\frac{1}{N}\sum_{j}\delta(z-y^{(j)}).\label{rho}
\ee
Since $C$ is analytic, singularities of the free energy are 
characterized by the DOZ only, which motivates us to investigate the DOZ.

The delta function can be rewritten as 
\be
\delta(y)=\frac{1}{2\pi}\lb \Part{}{y_1}{2}+\Part{}{y_2}{2} \rb \ln |y|,
\ee
which is the same relation as the one between a point charge and 
an electrostatic potential in electrostatics in two dimension. 
This relation leads to 
\be 
\rho(y_1,y_2)=
\frac{1}{2\pi}\lb \Part{}{y_1}{2}+\Part{}{y_2}{2} \rb\frac{1}{N}\ln |Z(y)|
\equiv \frac{1}{2\pi}\lb \Part{}{y_1}{2}+\Part{}{y_2}{2} \rb g(y).
\label{g}
\ee
This formula becomes a base of the following discussion. 

Note that \eref{g} is the Poisson equation, 
which means that the relation between the generating function $g(y)$ 
and the DOZ $\rho(y)$ is compared to the one between 
the electrostatic potential and the charge density in two dimension. 
The one-dimensionally-distributed charge density is evaluated 
from the discontinuity of the electric field. 
Similarly, the one-dimensionally-distributed DOZ appearing on 
phase boundaries can be assessed by the difference 
of first derivatives of $g(y)$ in the adjacent phases. 
Based on this analogy, we can derive the following formula 
for the one-dimensional DOZ $\rho_{\rm 1d}(y_1,y_2)$ 
on a phase boundary represented by $b(y_1,y_2)=0$:
\be
\rho_{\rm 1d}(y_1,y_2)=
\frac{1}{2\pi}
\lbb
\lb
\Part{g_{1}}{y_1}{}-\Part{g_{2}}{y_1}{}
\rb
\Part{b}{y_1}{}
+
\lb
\Part{g_{1}}{y_2}{}-\Part{g_{2}}{y_2}{}
\rb
\Part{b}{y_2}{}
\rbb
\delta(b),
\label{rho_1d}
\ee
where $g_1(y)$ and $g_2(y)$ are the generating functions 
of the adjacent phases. 
The function $b$ is defined such that $b$ is positive (negative) 
in the phase 1 (2) and becomes zero on the boundary given by $g_1=g_2$.
In most cases, it can be chosen as $b=g_1-g_2$.  

\subsection{Zeros of random systems and the replica method}
\label{sec:zeros by replica}

For random systems such as SGs, the DOZ fluctuates from sample to sample. In the thermodynamic limit, we can expect that the typical DOZ converges to the averaged one. This requires to take a difficult average of the logarithm as $[\ln |Z|]$, where the brackets $[(\cdots)]$ denote the average over the quenched randomness. The replica method bypasses this problem by using the identity
\be
g(y)= \frac{1}{N}[\ln |Z(y)|]
=\lim_{n\to 0}\frac{1}{2nN}\ln[|Z(y)|^{2n}]
\equiv \lim_{n\to 0}\frac{1}{2n}\phi(y,n).
\label{phi}
\ee
Once we obtain $\phi(y,n)$, we can calculate the DOZ from $\phi(y,n)$ through \eref{g} and \eref{phi}. Unfortunately, it is still difficult to treat the $n$th power for real $n$. To avoid this difficulty, we first assume that the exponent $n$ is an integer and evaluate $[|Z|^{2n}]$ in that condition. After that, we take the limit $n\to 0$ by utilizing the analytic continuation from integer to real. 

This standard prescription of the replica method has some delicate problems in taking $n \to 0$ limit. In some cases, a naive analytic continuation (RS solution) leads to an incorrect result, and the RSB solution is required. The RSB takes the rugged landscape of the free energy into account, which is essential to consider SG systems. In the present formulation, the RSB is implemented as an ansatz in the overlap matrix among $n$ replicas as usual. In our formulation to calculate $[|Z|^{2n}]=[(ZZ^*)^{n}]$, we have three types of overlaps: the usual overlap $\{q\}$ among $n$ replicas of $Z$, $\{q'\}$ among $n$ replicas of $Z^*$, and the inter-overlap $\{\tilde{q}\}$ between $n$ replicas of $Z$ and those of $Z^*$. Hence, we need some modifications in the RS and RSB ansatz to treat this extended overlap matrix. The detailed discussion about this point is presented in the next section after constructing the replica solution of the spherical SG.

Before closing this section, we mention a physical consequence of the inter-overlap $\{\tilde{q}\}$. If $\{\tilde{q}\}$ vanishes, the generating function decouples as $g(y)=[\ln Z+\ln Z^*]/N$ and the DOZ inevitably vanishes. However, the reverse is not necessarily true. Even when the inter-overlap takes a finite value, the DOZ can become zero. An actual example is shown in section~\ref{sec:diagrams}. We also mention that the replica method to calculate the zeros is very similar to that to find the chaos effect~\cite{Franz:95,Rizzo:01,Rizzo:02,Rizzo:03,Franz:92}. In both calculations, the replica space is doubled to find the nontrivial effects of inter-overlaps.

\section{Replica analysis of zeros of the spherical spin glass}
\label{sec:Replica}

The Hamiltonian of the $p$-body interacting spherical SG is given by
\be
\mathcal{H}=-\sum_{i_1<\cdots<i_p}J_{i_1 \cdots i_p}S_{i_1}\cdots S_{i_p},
\ee
where the spin $S_i$ takes continuous values under 
the spherical constraint $\sum_{i}S_i^2=N$, 
and the interaction $J_{i_1\cdots i_{p}}$ is drawn from 
Gaussian with the variance $J^2p!/2N^{p-1}$ 
\be
{\rm Prob}(J_{i_1\cdots i_p})
=\sqrt{\frac{N^{p-1}}{\pi J^2 p!}}
\exp\lb -\frac{N^{p-1}}{J^2p!}J_{i_1\cdots i_p}^2\rb.
\label{Jij}
\ee
In this paper, we also treat the multiple $(p+r)$-body interacting case, 
but below we explain our formulation on the single $p$-body interacting case. 
This is for the simplicity of the notation, 
but the analysis of the multiple-body case is essentially the same 
as the single one and hence the generalization is straightforward.

To assess the zeros in the complex temperature plane, we calculate $[|Z|^{2n}]$ 
by using the replica method as noted in the previous section. 
We can write $[|Z|^{2n}]=[Z^{n}(Z^{*})^n]$ 
under the assumption $n\in \mathbb{N}$ as follows:
\be
\hspace*{-1.0cm}
[|Z|^{2n}]&=&
\Tr 
\left[
 \exp 
  \lbb
\sum_{i_1<\cdots<i_p}J_{i_1 \cdots i_p} \lb 
  \beta \sum_{a=1}^{n}S_{i_1}^{a} \cdots S_{i_p}^{a}
+
  \beta^* \sum_{a=1}^{n}S_{i_1}^{'a} \cdots S_{i_p}^{'a}
\rb  
\rbb
\right]
\no \\ 
&&\times \prod_{a=1}^{n}
 \delta\lb \sum_{i=1}^{N}(S_{i}^{a})^2-N \rb 
 \delta\lb \sum_{i=1}^{N}(S_{i}^{'a})^2-N \rb,
\label{Z1}
\ee
where $\Tr$ means the integration over all the spin variables. 
The spherical constraint is expressed in the delta functions. 
The average $[(\cdots)]$ can be easily performed 
\be
\left[
 \exp 
  \lbb
\sum_{i_1<\cdots<i_p}J_{i_1 \cdots i_p} \lb 
  \beta \sum_{a=1}^{n}S_{i_1}^{a} \cdots S_{i_p}^{a}
+
  \beta^* \sum_{a=1}^{n}S_{i_1}^{'a} \cdots S_{i_p}^{'a}
\rb  
\rbb
\right]
\no \\ 
=\exp N \lbb 
\frac{\beta^2J^2}{4}\sum_{a,b}q_{ab}^p
+
\frac{(\beta^* )^2J^2}{4}\sum_{a,b}(q_{ab}^{'})^p
+
\frac{|\beta|^2J^2}{2}\sum_{a,b}\tilde{q}_{ab}^p 
\rbb, 
\label{config}
\ee
where we put 
\be
q_{ab}=\frac{1}{N}\sum_{i=1}^{N}S_i^{a}S_i^{b},\quad
q_{ab}'=\frac{1}{N}\sum_{i=1}^{N}S_i^{'a}S_i^{'b},\quad
\tilde{q}_{ab}=\frac{1}{N}\sum_{i=1}^{N}S_i^{a}S_i^{'b}.
\label{overlap}
\ee
Note that subleading terms are omitted by using the following relation
\be
\frac{p!}{N^{p}}\sum_{i_1<\cdots<i_p}(S_{i_1}^{a}S_{i_1}^{b})
\cdots (S_{i_p}^aS_{i_p}^b)
=\lb \frac{1}{N}\sum_{i=1}^NS_{i}^aS_{i}^b\rb^p+O(N^{-1}).
\ee
Let us express the relation \eref{overlap} by delta functions and 
introduce the overlaps $\{q_{ab},q'_{ab},\tilde{q}_{ab}\}$ 
as integration variables into \eref{config}. 
Besides, we also rewrite the delta functions by the Fourier expressions as
\be
\delta \lb\sum_{i=1}^{N}S_i^{a}S_i^{b}-N q_{ab} \rb
=\int d\chi_{ab}\exp \lbb A\chi_{ab}\lb 
\sum_{i=1}^{N}S_i^{a}S_i^{b}-N q_{ab}\rb \rbb.
\ee
The factor $A$ is arbitrary and chosen to make 
the following calculations simple. Hence, \eref{Z1} reads 
\be
\hspace{-2.4cm}[|Z|^{2n}] =
\int \lb \prod_{a,b}dq_{ab}dq'_{ab}d\tilde{q}_{ab} 
d\chi_{ab}d\chi'_{ab}d\tilde{\chi}_{ab}\rb
\exp N \Biggl\{ 
\frac{\beta^2J^2}{4}\sum_{a,b}q_{ab}^p
+
\frac{(\beta^* )^2J^2}{4}\sum_{a,b}(q_{ab}')^p
\no \\ \hspace{-0.5cm}
+
\frac{|\beta|^2J^2}{2}\sum_{a,b}\tilde{q}_{ab}^p 
-\frac{1}{2}\sum_{a,b}
\lb 
\chi_{ab}q_{ab}+\chi'_{ab}q'_{ab}+2\tilde{\chi}_{ab}\tilde{q}_{ab}
\rb
+\ln \Tr e^{L}
\Biggr\},
\label{Z2}
\ee
where
\be
 L&=&\frac{1}{2} \sum_{a,b}\lb\chi_{ab}S^{a}S^{b}
+\chi'_{ab}S^{'a}S^{'b}+2\tilde{\chi}_{ab}S^{a}S^{'b} \rb \no\\
 &=&
\frac{1}{2}
\left(\begin{array}{cc}
	S^{\rm T} & S^{'{\rm T}}
\end{array}\right)
\left(
\begin{array}{cc}
	X & \tilde{X} \\
	\tilde{X}^{\rm T} & X' 
\end{array}
 \right)
\left(
\begin{array}{c}
	S \\
	S' 
\end{array}
 \right). 
\ee
The matrix $X$ has the element $X_{ab}=\chi_{ab}$, 
and $X'$ and $\tilde{X}$ are defined in the same way. 
The matrix $\tilde{X}^{\rm T}$ represents the transposition of $\tilde{X}$. 
The spherical constraint is also expressed by the Fourier expression and 
is absorbed into the diagonal part of $X$ and $X'$. 
The factor $L$ is quadratic and the spin integration can be easily performed. 
The result is
\be 
\ln \Tr e^{L}=\frac{1}{2} \Tr \ln \lbb -2\pi 
\left(
\begin{array}{cc}
	X & \tilde{X} \\
	\tilde{X}^{\rm T} & X' 
\end{array}
 \right)^{-1}  \rbb.
\label{Tr}
\ee
Note that $\Tr$ in the right-hand side, and henceforth, 
denotes the trace of the matrix. 
In the thermodynamic limit, we can use the saddle-point method 
to evaluate \eref{Z2}. 
Substituting \eref{Tr} into \eref{Z2} and taking the saddle-point condition 
with respect to $X$, $X'$ and $\tilde{X}$, we find
\be
\left(
\begin{array}{cc}
	X & \tilde{X} \\
	\tilde{X}^{\rm T} & X' 
\end{array}
 \right)^{-1}  
=
-
\left(
\begin{array}{cc}
	Q & \tilde{Q} \\
	\tilde{Q}^{\rm T} & Q' 
\end{array}
 \right)\equiv -W_{Q}.
\label{s.p.X}
\ee
The matrix $Q$ is defined as $Q_{ab}=q_{ab}$, 
and $Q'$ and $\tilde{Q}$ are defined in the same way. 
Note that the real part of $W_Q$ should be positive 
for the convergence of the spin trace performed in \eref{Tr}. 
Summarizing the above transformations, 
we get $\phi(\beta,n)=\ln [|Z|^{2n}]/N$ as 
\be
\phi(\beta,n) &=& \frac{J^2}{4}\sum_{a,b}
\lb \beta^2 q_{ab}^p+(\beta^{*})^2 (q'_{ab})^p+2|\beta|^2 q_{ab}^p \rb \no\\
 && +\frac{1}{2}\Tr \ln W_{Q}+n(1+\ln 2 \pi). 
\label{phi-sp}
\ee
For further calculations, we need some ansatz with respect to $W_{Q}$. 
Note that the diagonal part of $W_{Q}$, $q_{aa}$ and $q_{aa}'$, 
is equal to unity $q_{aa}=q_{aa}'=1$. 
This is the consequence of the spherical constraint, 
which can be easily seen in \eref{overlap}.
We also have the condition that 
the matrix $W_Q$ is symmetric as $W_Q=W_Q^{\rm T}$.

\subsection{RS ansatz}
\label{sec:RS}

We start from the simplest RS case. 
Due to the additional overlaps coming from the complex parameter, 
even the RS solution has a nontrivial form. 
We first illustrate this point. 

Let us refer to the analysis of the REM given in~\cite{Ogure:04}. 
In that analysis, there are two types of RS solutions even in 
the real-parameter case. 
They are distinguished by the way of partitioning 
the $n$ replicas into the spin states. 
In one way, all the replicas are distributed into different states, 
and in the other one all the replicas are in a single state. 
Fortunately, these solutions are parameterized by 
a single overlap $q$ and we need not to distinguish them 
in the form of the overlap matrix. 
The first way gives the paramagnetic solution $q=0$ and 
the other yields the SG solution $q=1$.

\begin{figure}[htbp]
\begin{center}
\includegraphics[width=0.75\columnwidth]{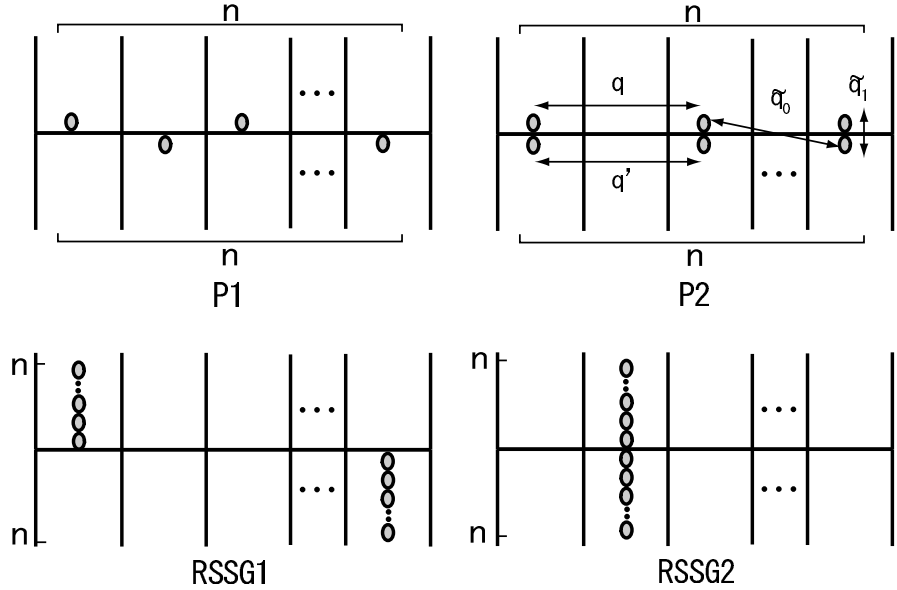}
\caption{Four types of RS solutions. 
Each ball represents a replica and each box a pure state. 
The upper half of the boxes corresponds to the phase space of 
the original partition function $Z$, and the lower one is 
the counterpart of $Z^*$. 
We assume that the number of pure states is larger than $n \in \mathbb{N}$.}
\label{fig:q-RS}
\end{center}
\end{figure}

To construct the correct RS form of the overlap matrix 
in the complex-parameter case, 
we need to generalize the above replica-partitioning way. 
Let us extend the interpretation of the ``state'' into 
the pure state\footnote{This interpretation includes 
the REM case, since a pure state of the REM coincides with a spin state.}. 
We assume that either of the above two-types solutions is realized 
if we focus only on the original $n$ replicas $Z^n$, 
or only on its conjugate $(Z^*)^n$. 
Considering the symmetry between $Z^n$ and $(Z^*)^n$, 
we obtain four types of RS solutions, 
which are schematically expressed in figure~\ref{fig:q-RS}.
Then, under this ansatz, the overlap matrices can be parameterized as
\be
\hspace*{-1.0cm}
Q=
\lb
\begin{array}{ccc}
  1 &  & q \\
   & \ddots &  \\
  q &  & 1 
\end{array}	      
\rb
,\quad
Q'=
\lb
\begin{array}{ccc}
  1 &  & q' \\
   & \ddots &  \\
  q' &  & 1 
\end{array}	      
\rb
,\quad
\tilde{Q}=
\lb
\begin{array}{ccc}
  \tilde{q}_1 &  & \tilde{q}_0 \\
   & \ddots &  \\
  \tilde{q}_0 &  & \tilde{q}_1 
\end{array}	      
\rb\label{RS}.
\ee
In each matrix, all the off-diagonal elements are filled in  
by the same parameter.
We note that any permutation of $\tilde{Q}$ also becomes a solution. 
This is because the indices of the original replicas $Z^n$ 
and those of the conjugate ones $(Z^*)^n$ can be independently chosen. 
We here present the simplest form. 
This solution includes that of REM with complex parameters~\cite{Takahashi:11}.

Two parameters of $\tilde{Q}$, $\tilde{q}_1$ and $\tilde{q}_0$, 
are needed to express the P2 solution and are the consequence of 
introduction of the complex parameter. 
Actually, the P2 solution gives two-dimensional distribution of zeros, 
which is a property distinguished from the other RS phases 
being characterized by $\tilde{q}_1=\tilde{q}_0$. 

Using this RS ansatz, we next calculate $\phi(\beta,n)$. 
We can write the first term in \eref{phi-sp}
\be
\sum_{a,b}q_{ab}^p=n+n(n-1)q^p,
\ee
and the second and third ones as well. 
To calculate the $\Tr$ term in \eref{phi-sp}, 
we need to diagonalize $W_{Q}$. 
The calculations are simple and omitted here. 
The result is
\be
\Tr{}\ln W_Q=\ln \lambda_{1}+(n-1)\ln\lambda_{0},
\ee
where
\be
&&\lambda_{1}=(1+(n-1)q)(1+(n-1)q')-(\tilde{q}_1+(n-1)\tilde{q}_0)^2,\\
&&\lambda_{0}=(1-q)(1-q')-(\tilde{q}_1-\tilde{q}_0)^2.
\ee
Substituting these terms into \eref{phi-sp}, we get
\be
\phi_{\rm RS}(\beta,n) &=& n\frac{J^2}{4}
\Biggl\{
\beta^2\lb 1+(n-1)q^{p} \rb
+
(\beta^*)^2 \lb 1+(n-1)(q')^{p} \rb
\no\\
& & +2|\beta|^2 \lb \tilde{q}_1^{p}+(n-1)\tilde{q}_0^{p} \rb
\Biggr\}
\no \\
& & +\frac{1}{2}\lb \ln\lambda_{1}+(n-1)\ln \lambda_0 \rb
+n\lbb 1+\ln(2\pi) \rbb.
\ee
In the limit $n\to 0$, we have
\be
\lambda_{1} \sim \lambda_{0}+n\lbb 
q(1-q')+q'(1-q)-2\tilde{q}_{0}(\tilde{q}_1-\tilde{q}_0)
\rbb.
\ee
Then, we get $g_{\rm RS}(\beta)=\lim_{n\to 0}\phi_{\rm RS}(\beta,n)/2n$ as
\be
g_{\rm RS}(\beta) &=& \frac{J^2}{8}
\Biggl\{
\beta^2\lb 1-q^{p} \rb
+
(\beta^*)^2 \lb 1-(q')^{p} \rb
+
2|\beta|^2 \lb \tilde{q}_1^{p}-\tilde{q}_0^{p} \rb
\Biggr\}
\no \\
&&+\frac{1}{4}
\lbb
\ln \lambda_0 
+\frac{q(1-q')+q'(1-q)-2\tilde{q}_{0}(\tilde{q}_1-\tilde{q}_0)}{\lambda_0}
\rbb
\no\\
&& +\frac{1}{2}\lbb 1+\ln( 2\pi )\rbb.
\label{g-RS}
\ee
The saddle-point conditions yield
\be
&&\hspace{-2.cm}\mu_p q^{p-1}-\frac{1}{\lambda_{0}^2}
\lbb q(1-q')^2+(\tilde{q}_1-\tilde{q}_0)\lb
	-2\tilde{q}_0+q'(\tilde{q}_1+\tilde{q}_0) \rb \rbb=0
,
\label{EOS-RS-q}
\\
&&\hspace{-2.cm}\mu_p^* q'^{p-1}-\frac{1}{\lambda_{0}^2}
\lbb q'(1-q)^2+(\tilde{q}_1-\tilde{q}_0)\lb 
	-2\tilde{q}_0+q(\tilde{q}_1+\tilde{q}_0) \rb \rbb=0
\label{EOS-RS-q'}
,\\
&&\hspace{-2.cm}|\mu_p| \tilde{q}_{1}^{p-1}-\frac{1}{\lambda_{0}^2}
\Biggl\{ \tilde{q}_1\lb (1-q)(1-q')+(\tilde{q}_1-\tilde{q}_0)^2 \rb
\no \\
&&\hspace{0cm}
-(\tilde{q}_1-\tilde{q}_0)\lb q+q'-2qq' \rb -2(\tilde{q}_1-\tilde{q}_0)^3
\Biggr\}
=0
\label{EOS-RS-qtil1}
,\\
&&\hspace{-2.cm}|\mu_p| \tilde{q}_{0}^{p-1}-\frac{1}{\lambda_{0}^2}
\lbb \tilde{q}_0\lb (1-q)(1-q')+(\tilde{q}_1-\tilde{q}_0)^2 \rb
-(\tilde{q}_1-\tilde{q}_0)\lb q+q'-2qq' \rb \rbb=0
,\label{EOS-RS-qtil0}
\ee
where we put $\mu_p=p\beta^2 J^2/2$.

\subsubsection{Remarks for RS solutions}
\label{sec:RSremarks}

In figure~\ref{fig:q-RS}, we assumed that a set of spin configurations 
consisting a pure state with the weight $e^{-\beta \mathcal{H}(\bm{S})}/Z$ also consists a pure state with the conjugate weight 
$(e^{-\beta \mathcal{H}(\bm{S}')})^*/Z^*$. 
This can be accepted by considering that the two weights of 
an identical spin configuration $\bm{S}=\bm{S}'$ take 
the same absolute value 
$|e^{-\beta \mathcal{H}(\bm{S})}/Z|=|(e^{-\beta \mathcal{H}(\bm{S})})^*/Z^*|$, 
which implies that the support of each weight becomes identical. 
This indicates that the above assumption holds, 
since each pure state can be regarded as a support of the Boltzmann weight.

Each pure state $a$ has its own partition function $Z_{a}$, 
and the total partition function is given by $Z=\sum_{a}Z_{a}$. 
Using this notation and the standard description of 
pure states~\cite{SPIN} and focusing on the P2 solution as an example, we can write the physical significance of the overlap $q$ as
\be
q=\sum_{a \neq b}w_a w_b
\sum_{i}\frac{1}{N}
\frac{\Tr S_{i}e^{-\beta \mathcal{H}(\bm{S}) }\delta_{a}(\bm{S})}{Z_{a}}
\frac{\Tr S_{i}e^{-\beta \mathcal{H}(\bm{S}) }\delta_{b}(\bm{S})}{Z_{b}},
\label{q-pure}
\ee
where we put $w_a=Z_a/Z$ and introduce an indicator function 
$\delta_{a}(\bm{S})$ which is defined as $\delta_{a}(\bm{S})=1$ if 
$\bm{S}$ belongs to the pure state $a$ and $\delta_{a}(\bm{S})=0$ otherwise. 
Similarly, the conjugate overlap $q'$ becomes 
\be
q'=\sum_{a \neq b}w_a^* w_b^*
\sum_{i}\frac{1}{N}
\frac{\Tr S_{i}(e^{-\beta \mathcal{H}(\bm{S}) })^* \delta_{a}(\bm{S})}{Z_{a}^*}
\frac{\Tr S_{i}(e^{-\beta \mathcal{H}(\bm{S}) })^* \delta_{b}(\bm{S})}{Z_{b}^*},
\label{q'-pure}
\ee
and the inter-overlaps $\tilde{q}_1$ and $\tilde{q}_0$ are expressed as
\be
\tilde{q}_1=\sum_{a}w_a w_a^*
\sum_{i}\frac{1}{N}
\frac{\Tr S_{i}e^{-\beta \mathcal{H}(\bm{S})} \delta_{a}(\bm{S})}{Z_{a}}
\frac{\Tr S_{i}(e^{-\beta \mathcal{H}(\bm{S})})^* \delta_{a}(\bm{S})}{Z_{a}^*},
\label{qtil1-pure}
\\
\tilde{q}_0=\sum_{a \neq b}w_a w_b^*
\sum_{i}\frac{1}{N}
\frac{\Tr S_{i}e^{-\beta \mathcal{H}(\bm{S})} \delta_{a}(\bm{S})}{Z_{a}}
\frac{\Tr S_{i}(e^{-\beta \mathcal{H}(\bm{S})})^* \delta_{b}(\bm{S})}{Z_{b}^*}.
\label{qtil0-pure}
\ee
Combining \eref{q-pure}-\eref{qtil0-pure} and the fact that 
the pure states are common for $Z$ and $Z^*$, 
we can derive the constraints for the overlaps as 
\be
q'=q^*, \quad
\tilde{q_1}=\tilde{q_1}^*, \quad
\tilde{q_0}=\tilde{q_0}^*.
\label{q-constraint}
\ee 
We can easily find that \eref{EOS-RS-q}-\eref{EOS-RS-qtil0} actually 
have the solutions satisfying these constraints. For the rest of this paper, 
we assume \eref{q-constraint} from the beginning of analyses.

\subsection{1RSB}
\label{sec:1RSB}

\begin{figure}[htbp]
\begin{center}
\includegraphics[width=0.75\columnwidth]{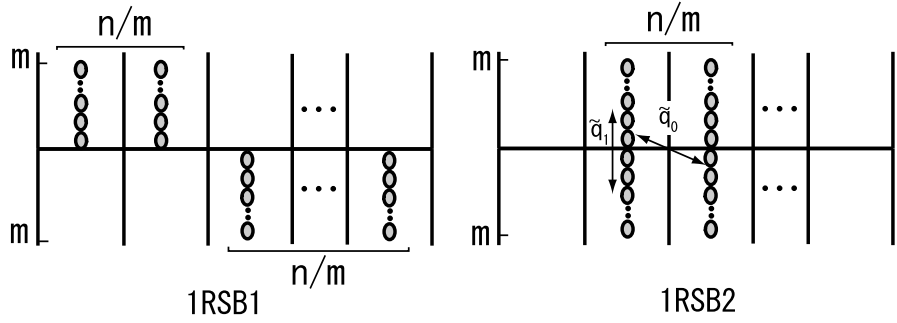}
\caption{Two types of 1RSB solutions.}
\label{fig:q-1RSB}
\end{center}
\end{figure}

Next, we derive the 1RSB solutions of \eref{phi-sp}. 
In a similar way as the RS case, 
we have two types of solutions represented in figure~\ref{fig:q-1RSB}.
The corresponding overlap matrices are given by, e.g. for the $(n,m)=(4,2)$ case,
\be
\hspace{-2cm}
Q=
\lb
\begin{array}{cc|cc}
  1   & q_1 & q_0& q_0 \\
  q_1 & 1   & q_0& q_0 \\
\hline
  q_0 & q_0 & 1  & q_1 \\
  q_0 & q_0 & q_1& 1  
\end{array}	      
\rb
,
\quad
Q'=
\lb
\begin{array}{cc|cc}
  1 & q'_1 & q'_0 & q'_0 \\
  q'_1 & 1 & q'_0 & q'_0 \\
\hline
  q'_0 & q'_0 & 1 & q'_1 \\
  q'_0 & q'_0 & q'_1 & 1  
\end{array}	      
\rb
,
\quad
\tilde{Q}=
\lb
\begin{array}{cc|cc}
  \tilde{q}_1 & \tilde{q}_1 & \tilde{q}_0& \tilde{q}_0 \\
  \tilde{q}_1 & \tilde{q}_1 & \tilde{q}_0& \tilde{q}_0 \\
\hline
  \tilde{q}_0 & \tilde{q}_0 & \tilde{q}_1 & \tilde{q}_1 \\
  \tilde{q}_0 & \tilde{q}_0 & \tilde{q}_1 & \tilde{q}_1  
\end{array}	      
\rb.\no\\
\label{q-1RSB}
\ee
After some calculations, we get
\be
&&
\hspace{-2.cm}
\phi_{\rm 1RSB}(\beta,n,m)=
\frac{n}{2p}\Biggl\{ 
\mu_p \lb 1+(m-1)q_{1}^p+(n-m)q_{0}^p \rb 
\no \\
&&
\hspace{-2.cm}
+\mu_p^* \lb 1+(m-1)(q'_{1})^p+(n-m)(q'_{0})^p \rb
+2|\mu_p| \lb m\tilde{q}_{1}^p+(n-m)\tilde{q}_{0}^p \rb
\Biggr\} 
\no \\
&&
\hspace{-2.cm}
+\frac{1}{2}\lb\ln \eta_2+\lb \frac{n}{m}-1 \rb\ln \eta_1
+\frac{n}{m}(m-1)\ln \eta_0 \rb+n\lbb 1+\ln(2\pi) \rbb,
\ee
where
\be
&&
\hspace{-2.5cm}
\eta_{0}=(1-q_1)(1-q'_1),\\
&&
\hspace{-2.5cm}
\eta_{1}=(1+(m-1)q_1-mq_0)(1+(m-1)q'_1-mq'_0)
-m^2(\tilde{q}_1-\tilde{q}_0)^2,\\
&&
\hspace{-2.5cm}
\eta_2=(1+(m-1)q_1+(n-m)q_0)(1+(m-1)q'_1+(n-m)q'_0)
-(m\tilde{q}_1+(n-m)\tilde{q}_0)^2 
\no \\
&&
\hspace{-2.5cm}
\sim \eta_1+n
\lbb
q_0\lb 1+(m-1)q'_1-mq'_0 \rb+q'_0\lb 1+(m-1)q_1-mq_0 \rb
-2m \tilde{q}_0 \lb \tilde{q}_1-\tilde{q}_0 \rb
\rbb.
\ee
For simplicity, let us hereafter assume the condition 
$q_{0}=q'_{0}=\tilde{q}_{0}=0$, which is expected to be satisfied 
due to the spin-reversal symmetry. 
Under this condition, we obtain 
$g_{\rm 1RSB}(\beta,m)=\lim_{n\to 0}\phi_{\rm 1RSB}(\beta,n,m)/2n$ as 
\be
&&
\hspace{-1.5cm}
g_{\rm 1RSB}(\beta,m)=\frac{1}{4p}\Biggl\{ 
\mu_p \lb 1+(m-1)q_{1}^p\rb 
+\mu_p^* \lb 1+(m-1)(q'_{1})^p\rb +2m|\mu_p| \tilde{q}_{1}^p
\Biggr\}
\no \\
&&
\hspace{-0.cm}
+\frac{1}{4m}\left\{ \ln \eta_1+(m-1)\ln \eta_0 \right\}
+\frac{1}{2}\lbb 1+\ln (2\pi) \rbb.
\label{g-1RSB}
\ee
Taking the variation with respect to $q_1$, $\tilde{q}_1$ and $m$, we get
\be
 \mu_p q_{1}^{p-1}+\frac{1}{m}
\lbb 
\frac{1+(m-1)q'_1}{\eta_1}-\frac{1-q'_1}{\eta_0}
\rbb =0,
\label{EOS-1RSB-q1}\\
 |\mu_p| \tilde{q}_{1}^{p-1}-
\frac{\tilde{q}_1}{\eta_1}
=0,
\label{EOS-1RSB-qtil1}
\\
\frac{1}{4p}
\lbb 
\mu_p q_{1}^p+\mu_p^* (q'_{1})^p+2|\mu_p|\tilde{q}_1^p
\rbb
-\frac{1}{4m^2}\left\{ \ln \eta_1+(m-1)\ln \eta_0 \right\}
\no 
\\
+\frac{1}{4m}\left\{ 
\frac{q_1(1+(m-1)q'_1)+q'_1(1+(m-1)q_1)-2m\tilde{q}_1^2}{\eta_1}
+\ln \eta_0
\right\}=0,
\label{EOS-1RSB-m}
\ee
where the saddle-point condition with respect to $q'_1$ 
is omitted since it gives the complex conjugate of $q_1$, 
the reason of which is the same as explained in section~\ref{sec:RSremarks}.

\section{Phase diagrams and DOZ}
\label{sec:diagrams}

\subsection{$p=2$ case}

\begin{figure}[htbp]
\begin{center}
\includegraphics[height=0.39\columnwidth,width=0.455\columnwidth]{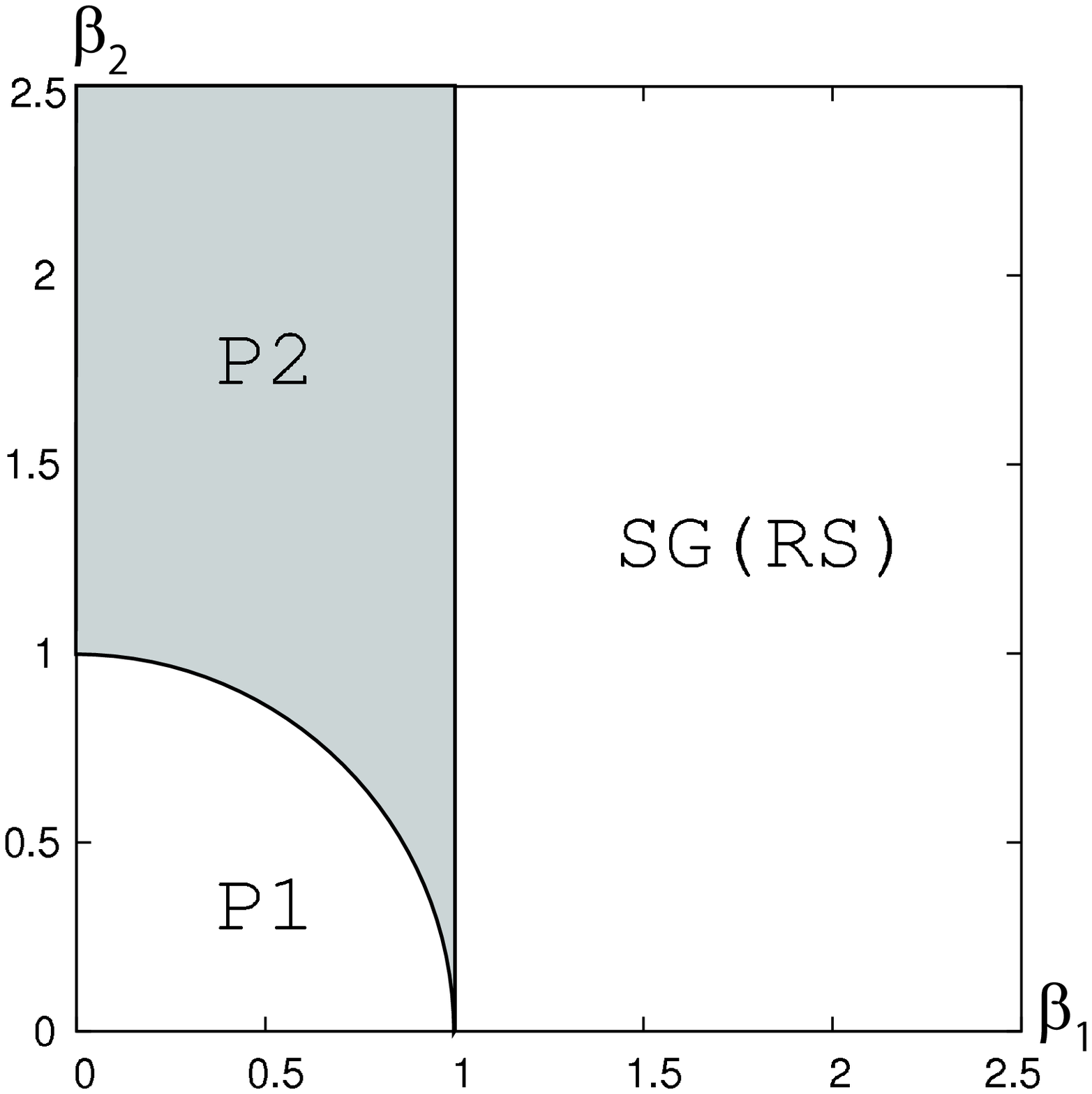}
\hspace{4mm}
\includegraphics[height=0.4\columnwidth,width=0.46\columnwidth]{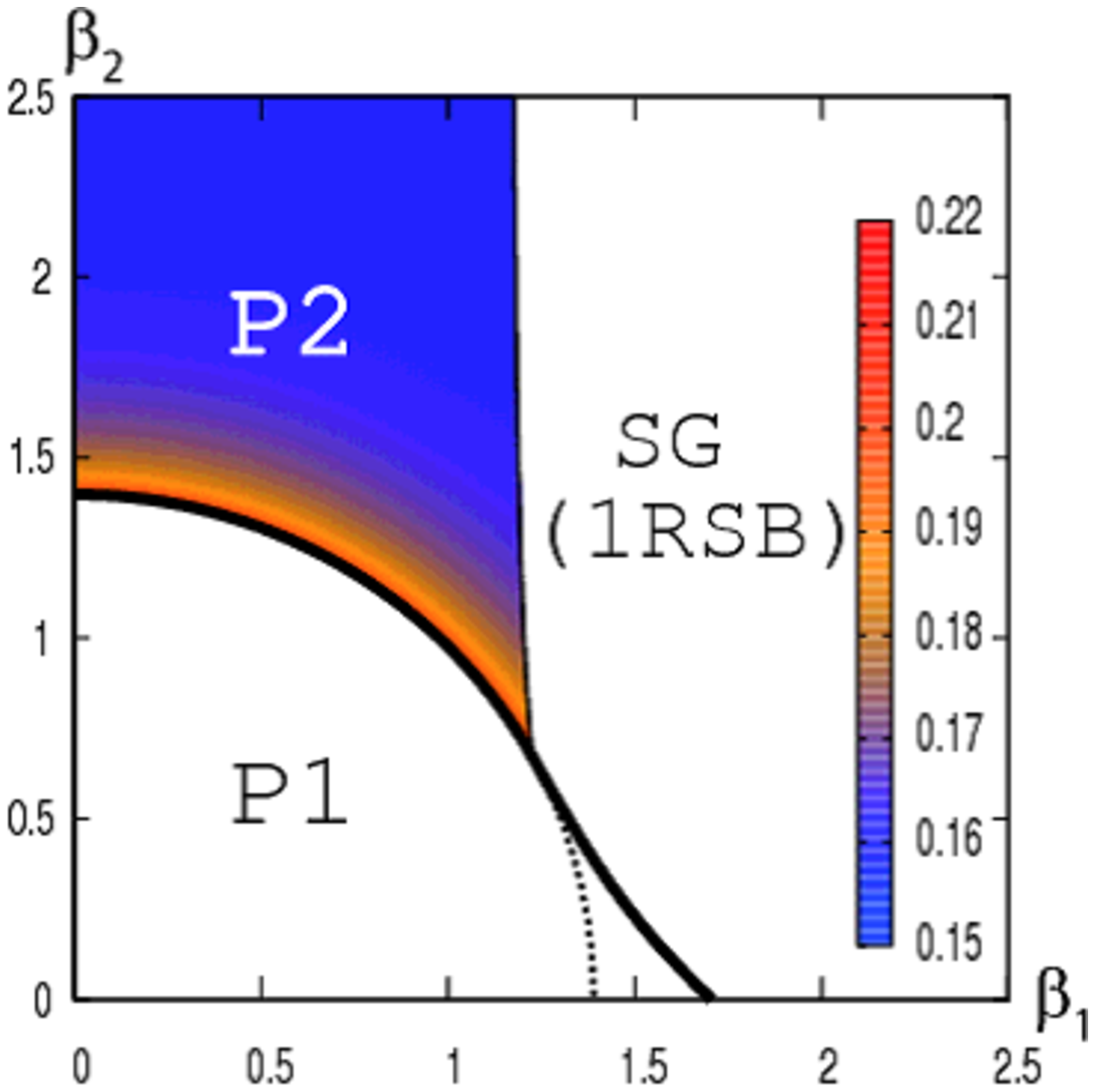}
\caption{Complex-temperature phase diagrams of the $p$-body spherical SG for $p=2$ (left) and $p=3$ (right) with $J=1$. The DOZ of the P2 phase for $p=2$ stays constant and is shaded, while that for $p=3$ is not constant and is coloured depending on the values. The SG and P1 phases have no DOZ in both the cases. On bold phase boundaries, the P2-P1 and P2-SG ones in the right panel, the one-dimensional DOZ becomes finite. The dashed line in the right panel is the extension of the boundary between P1 and P2 phases and is not the true phase boundary. }
\label{fig:PD}
\end{center}
\end{figure}

It is known that the RS solution is sufficient for the $p=2$ case. 
In this case, \eref{EOS-RS-q} and \eref{EOS-RS-qtil0} are 
low-degree polynomial equations and can be analytically solved. 
Based on the physical descriptions in figure~\ref{fig:q-RS}, 
we get the following three solutions: 

\begin{description}
\item[P1]{This solution is the usual paramagnetic solution 
$q=q'=\tilde{q}_1=\tilde{q}_0=0$. 
The generating function $g(\beta)$ becomes
\be
\hspace{-2cm}
g_{\rm P1}(\beta)=\frac{1}{4p}(\mu_p+\mu_p^*)
+\frac{1}{2}\lbb 1+\ln (2\pi)\rbb
=\frac{1}{4}(\beta_1^2-\beta_2^2)J^2+\frac{1}{2}\lbb 1+\ln (2\pi)\rbb,
\label{g_P1}
\ee
and the corresponding DOZ is $\rho_{\rm P1}=0$.}

\item[P2]{This solution is given by 
$q=q'=\tilde{q}_0=0$ and $\tilde{q}_1>0$. 
Assuming $q=q'=0$, we can easily solve \eref{EOS-RS-qtil1} as
\be
\tilde{q}_1^2=1-\frac{1}{|\beta|^2J^2}.
\ee
The inter-overlap $\tilde{q}_1$ should be real 
as explained in section~\ref{sec:RSremarks}, which means that 
this solution is valid only for $|\beta|J>1$. 
Substituting this solution into \eref{g-RS}, we get
\be
g_{\rm P2}(\beta)=\frac{\beta_1^2J^2}{2}-\frac{1}{4}\lbb 1
+ \ln (|\beta|^2J^2) \rbb+\frac{1}{2}\lbb 1+ \ln (2\pi) \rbb.
\label{g_P2}
\ee
The corresponding DOZ yields a finite value 
\be 
\rho_{\rm P2}=\frac{J^2}{2\pi},
\ee
which is the same value as the REM's one~\cite{Derrida:91,Takahashi:11}.
}

\item[RSSG]{We impose $q,q'\neq 0$ and $\tilde{q}_1=\tilde{q}_0=\tilde{q}$ 
in \eref{g-RS} and find that $g(\beta)$ does not depend on $\tilde{q}$. 
This is because the contribution of $\tilde{q}$ is 
proportional to $O(n^2)$ in \eref{phi-sp} and vanishes in the limit $n\to 0$. 
This means that we cannot distinguish the solutions RSSG1 and RSSG2 
in figure~\ref{fig:q-RS}, and hence we just call this RSSG. 
The solution of \eref{EOS-RS-q} and \eref{EOS-RS-q'} is given by
\be
q=1-\frac{1}{\sqrt{\beta^2}J},\quad q'=q^*.
\ee
A condition $\Re{q}>0$ required in taking the spin trace of \eref{Tr} 
leads to $q=1-1/\beta J$ for $\beta_1>0$ and 
$q=1+1/\beta J$ for $\beta_1<0$. 
The generating function then becomes
\be
g(\beta)=|\beta_1|J-\frac{1}{4}\lbb 3+\ln(|\beta|^2J^2) \rbb
+\frac{1}{2}\lbb 1+\ln (2\pi)\rbb.
\ee
We can easily find that the DOZ also vanishes 
as the P1 case, $\rho_{\rm RSSG}=0$.}
\end{description}
Summarizing these results, we can derive the phase diagram and 
show it in the left panel of figure~\ref{fig:PD}.
The phase boundaries are obtained by equating the generating functions 
$g(\beta)$ of the adjacent phases. 
Note that $g_{\rm P2}$ is always larger than $g_{\rm P1}$ 
and $g_{\rm RSSG}$ except for on the phase boundaries, which seemingly 
implies that the P2 solution dominates all the complex $\beta$ plane, 
if we blindly follow the saddle-point method. 
Obviously, this is incorrect. 
The constraint $\tilde{q}_1=\tilde{q}_1^*$ means 
the failure of the P2 solution in a region $|\beta|J<1$, 
which explains the emergence of the P1 solution. 
On the other hand, we have no direct reason 
to explain the phase transition from P2 to RSSG. 
We choose the RSSG phase in the region $\beta_1>1$ based on 
the physical appropriateness. 
This point will be clearer by considering the region $n>0$, 
but it is beyond our current purpose in this paper. 

The DOZ on the boundaries should be evaluated separately 
by using~\eref{rho_1d}. 
For the P1-P2 boundary, we can calculate the DOZ by 
putting $g_{1}=g_{P1}$, $g_{2}=g_{P2}$ and 
$b(\beta_1,\beta_2)=1-(\beta_1^2+\beta_2^2)J^2$ in~\eref{rho_1d}. 
Note that we cannot put $b=g_{P1}-g_{P2}$ in this case, 
since the P2 solution is always larger than or equal to 
the other solutions as explained in the above paragraph. 
Simple calculations show that the density vanishes. 
Similarly, we can evaluate the density on the P2-SG boundary 
and again it becomes zero.  

Based on the derived DOZ, we can directly calculate the specific heat 
$C(\beta)$ by the formula
\be
C(\beta)=-\int dz_1 dz_2 \rho(z_1,z_2)\frac{\beta^2}{(\beta-z)^2}.
\ee
We find that the result completely agrees 
with the known values~\cite{Kosterlitz:76}, which validates our calculation.

We also notice that the DOZ inevitably vanishes in the RSSG phases. 
This is because the absence of contributions from 
the inter-overlaps $\tilde{q}$, where the generating function $g(\beta)$ 
becomes analytic with respect to $\beta$. 
This consideration assures that there is no kind of transitions 
or chaos effects in the RSSG phase.

\subsection{$p=3$ case}
\label{sec:PDp=3}

We first derive the phase diagram on the complex-temperature plane.
Here, we investigate only the $p=3$ case, but the result is expected 
to be essentially common for $p>3$. 
In the following discussion, we treat the 1RSB2 solution described 
in \fref{fig:q-1RSB} as the correct SG solution in this case. 
We find that the other 1RSB solution, 
the 1RSB1 one with $\tilde{q}_1=\tilde{q}_0=0$ in \fref{fig:q-1RSB}, 
actually exists but it shows an unphysical behaviour. 
Hence we reject it. 

The solutions P1 and P2 are essentially the same as the $p=2$ case. 
The difference between $g_{\rm P1}$ and $g_{\rm P2}$ depends 
only on $|\beta|$, and the boundary between these phases becomes 
a circle whose radius is obtained by comparing the values of 
$g_{P1}$ and $g_{P2}$ with substitution of the solution 
of \eref{EOS-RS-qtil1} under the condition $q=q'=0$. 
The resultant radius becomes $|\beta|=\beta_p\approx 1.39884/J$. 

The SG phase of the $p=3$ case is known to be described 
by the 1RSB solution~\cite{Crisanti:92}. 
We derive the transition temperature $\beta_c$ from P1 to SG 
at the real axis in our formulation. 
The equations of state \eref{EOS-1RSB-q1}-\eref{EOS-1RSB-m} have 
some solutions even at $\beta_2=0$. 
A reasonable solution among them is obtained under the condition 
$q_1=q'_1=\tilde{q}_1$, since the usual overlap matrix with 
a real temperature is recovered by this condition. 
Actually from \eref{EOS-1RSB-q1}-\eref{EOS-1RSB-m}, 
we can derive the same equations of state as the one under 
the usual real-parameter case by putting $q_1=q'_1=\tilde{q}_1$ 
and assuming $\beta$ is real~\cite{Crisanti:92}, 
though the breaking parameter $m$ in the usual case is replaced by $2m$ 
in our formulation. 
This is natural since each replica is doubled to calculate 
$|Z|^{2n}=(ZZ^*)^n$ in our formulation. 
Hence, additionally assuming the condition $m=1/2$ 
(which corresponds to $m=1$ in the usual formulation), 
we can calculate the transition temperature $\beta_c$ 
by solving \eref{EOS-1RSB-q1} and \eref{EOS-1RSB-m}, 
which leads to the known value $\beta_c\approx 1.70633/J$~\cite{Crisanti:92}.

We can easily confirm that $g_{\rm 1RSB}$ with the condition $m=1$ 
accords with $g_{\rm P2}$. 
This implies that the boundary between P2 and SG phases is obtained 
by solving \eref{EOS-1RSB-q1}-\eref{EOS-1RSB-m} under the condition $m=1$, 
which is actually the case for the REM~\cite{Takahashi:11}. 
This can also be seen from that the 1RSB equation of 
$\tilde{q}_1$ \eref{EOS-1RSB-qtil1} coincides with 
that of P2 \eref{EOS-RS-qtil1} at $m=1$. 
Solving \eref{EOS-1RSB-q1}-\eref{EOS-1RSB-m} under the condition $m=1$ 
involves some technical difficulties, 
the details of which are given in \ref{sec:app1}. 

In the SG phase, we need to directly treat all the equations of state 
\eref{EOS-1RSB-q1}-\eref{EOS-1RSB-m} to calculate $g_{\rm 1RSB}$. 
The phase boundary between the SG and P1 phases is obtained by 
equating $g_{\rm P1}$ and $g_{\rm 1RSB}$. 
The technical difficulties to evaluate \eref{EOS-1RSB-q1}-\eref{EOS-1RSB-m} 
in this case are also summarized in \ref{sec:app1}. 

Summarizing the above points, we can write the phase diagram for the $p=3$ 
case and give it in the right panel of figure~\ref{fig:PD}. 
We can find some difference from the $p=2$ case in the shape of the diagram. 
The shape is actually related to the DOZ on the boundaries, 
which is explained below. 

\subsubsection{Density of zeros}

Here, we calculate the DOZ for the $p=3$ case. 

The P1 phase is the trivial case. 
We can easily confirm that the DOZ of P1 phase, $\rho_{\rm P1}$, 
is uniformly zero as for $p=2$. 

\begin{figure}[htbp]
\begin{center}
\includegraphics[width=0.45\columnwidth]{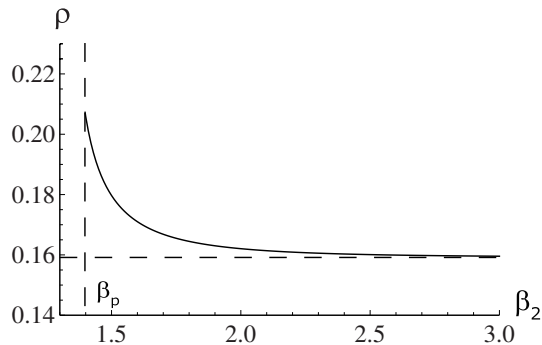}
 \caption{The DOZ on the imaginary axis $\beta_1=0$ for the $p=3$ case with $J=1$. The vertical dashed line denotes the transition point $\beta_2=\beta_p$ between the P1 and P2 phases and the horizontal one represents a constant $J^2/2\pi$ being the limiting value of $\rho_{\rm P2}$ in the limit $\beta_2\to \infty$.}
 \label{fig:rho-P2}
\end{center}
\end{figure}

The case of the P2 is more complicated. 
We can rewrite the generating function as 
$g_{\rm P2}=g_{\rm P1}+\Delta(\beta_1,\beta_2,\tilde{q}_1(\beta_1,\beta_2))$, 
where 
\be
\Delta(\beta_1,\beta_2,\tilde{q}_1(\beta_1,\beta_2))
=\frac{1}{4}(\beta_1^2+\beta_2^2)J^2\tilde{q}_1^{p}
+\frac{1}{4}\ln(1-\tilde{q}_1^{2}),
\label{Delta}
\ee
and the finite contribution to $\rho_{\rm P2}$ only comes from $\Delta$. 
Differentiating $\Delta$ with respect to $\beta_1$ twice, we get three terms
\be
\hspace{-2cm}
\lb \Part{\Delta}{\beta_1}{2}\rb_{\beta_2}=
\lb \Part{\Delta}{\beta_1}{2} \rb_{\beta_2,\tilde{q}_1}
+2\lb 
\frac{ \partial^2 \Delta }{ \partial\beta_1 \partial\tilde{q}_1 }
\rb_{\beta_2,\tilde{q}_1}
\lb
\Part{\tilde{q}_1}{\beta_1}{}
\rb_{\beta_2}
+
\lb 
\Part{\Delta}{\tilde{q}_1}{2}
\rb_{\beta_2,\tilde{q}_1}
\lb \Part{\tilde{q}_1}{\beta_1}{} \rb^2_{\beta_2},
\ee
where we omit a term being proportional to 
$\partial \Delta/\partial \tilde{q}_1$ since it vanishes 
due to the saddle-point condition. 
Subscripts of the brackets denote the fixed variables 
in taking the partial differentiation. 
Evaluating each term yields
\be
\lb \Part{\Delta}{\beta_1}{2} \rb_{\beta_2,\tilde{q}_1} 
 =\frac{1}{2}J^2\tilde{q}_1^p, \\
\lb \frac{ \partial^2 \Delta }{ \partial\beta_1 \partial\tilde{q}_1 } 
\rb_{\beta_2,\tilde{q}_1}
=\frac{p}{2} \beta_1 J^2 \tilde{q}_1^{p-1},\\
\lb \Part{\Delta}{\tilde{q}_1}{2} \rb_{\beta_2,\tilde{q}_1}
=\frac{p(p-1)}{4}(\beta_1^2+\beta_2^2)J^2\tilde{q}_1^{p-2}
-\frac{1}{2}\frac{1+\tilde{q}_1^2}{(1-\tilde{q}_1^2)^2},
\ee
and the factor $\partial \tilde{q}_1/\partial \beta_1$ 
can be calculated by differentiating \eref{EOS-RS-qtil1} 
with respect to $\beta_1$. 
The result is 
\be
\Part{\tilde{q}_1}{\beta_1}{} 
= p \beta_1 J^2 \tilde{q}_1^{p-2}
\left\{ \frac{(1-\tilde{q}_1^2)^2}{2\tilde{q}_1-|\mu_p|(p-2)
\tilde{q}_1^{p-3}(1-\tilde{q}_1^2)^2}\right\}
\equiv p\beta_1 J^2 \tilde{q}_1^{p-2}Y.
\ee
The counterpart with respect to $\beta_2$ can be obtained as well. 
Summing up both the contributions, we get
\be
\rho_{ {\rm P2} }&=&
\frac{J^2}{2\pi}\tilde{q}_1^{p}
+\frac{p^2(\beta_1^2+\beta_2^2)J^4\tilde{q}_1^{2p-3}Y}{2\pi}\no \\
&&+\frac{p^2(\beta_1^2+\beta_2^2)J^4\tilde{q}_1^{2p-4}}{4\pi}
\left\{|\mu_p|(p-1)\tilde{q}_1^{p-2}
-\frac{1+\tilde{q}_1^2}{(1-\tilde{q}_1^2)^2}\right\} Y^2.
\label{rho_P2}
\ee
We can easily evaluate this equation after solving \eref{EOS-RS-qtil1}. In figure~\ref{fig:PD}, $\rho_{{\rm P2}}$ is coloured on the phase diagram. We can find that the density tends to decrease as $\beta_2$ grows. To see the quantitative behavior more precisely, we plot the density on the imaginary axis in figure~\ref{fig:rho-P2}. The figure shows that the value of the density converges to a constant as $\beta_2$ grows, the limiting value of which is $J^2/2\pi$ as the $p=2$ case, which can be understood from \eref{EOS-RS-qtil1} and \eref{rho_P2}.

The DOZ of the SG phase can be assessed in a similar way to the P2 case, 
although it requires rather involved calculations 
due to the existence of four variational variables 
$q_1$, $q'_1$, $\tilde{q}_1$ and $m$. 
The resultant formula is not enlightening and we here omit it. 
Evaluating the DOZ through the formula, 
we find that there are no zeros in the SG phase\footnote{
We numerically evaluate the DOZ at several regions in the SG phase, 
such as around the P2-SG boundary and on a line $\beta_1=\beta_c$,
and confirm that the values are smaller than $10^{-12}$ at most.
This is in a margin of numerical errors of our calculation and 
we conclude that the DOZ is zero.}, 
which is the same as the REM. 
This implies the absence of any kind of phase transitions 
in the SG phase of this system. 

Next, we refer to the DOZ on the phase boundaries. 
First derivatives of $g(\beta)$ of adjacent phases are crucial as already noted in \eref{rho_1d}. 

On the P1-P2 boundary, choosing $g_1=g_{\rm P1}$, $g_{2}=g_{\rm P2}$ 
and $b=g_{\rm P1}-g_{\rm P2}$ in \eref{rho_1d}, 
we find the one-dimensional DOZ $\rho_{\rm P1-P2}$ as
\be
\rho_{\rm P1-P2}(\beta_1,\beta_2)=\frac{1}{4}J^4\tilde{q}_1^{2p}
\lb \beta_1^2+\beta_2^2 \rb
\delta\lb \Delta(\beta_1,\beta_2,\tilde{q}_{1})\rb,
\ee
where $\Delta$ is given in \eref{Delta}.
Clearly, this yields nonzero value, which is natural 
since the transition between the P1 and P2 phases is of first order 
in the $p=3$ case. 
Note that the derivative of $\tilde{q}_{1}$, which involves 
the factor $\partial g_{\rm P2}/\partial \tilde{q}_1$, does not appear 
in this formula, since $\partial g_{\rm P2}/\partial \tilde{q}_1$ 
vanishes due to the saddle-point condition.  

We can also confirm that the DOZ takes finite values 
on the P1-SG boundary by simple calculations. 
Since there are no zeros in both the phases, 
the Yang-Lee theorem, which proves no phase transition 
in a region without the DOZ, requires that 
the one-dimensional DOZ on the boundary cannot vanish.
This is in contrast to the $p=2$ case where the P2 phase intercepts 
the P1 and SG phases. 

Meanwhile, on the P2-SG boundary, the one-dimensional DOZ does not appear, 
since the condition $m=1$ at the boundary makes the first derivatives 
of $g_{\rm P2}$ and $g_{\rm 1RSB}$ identical. 
This observation implies that there is no one-dimensional density 
on the P2-SG boundary in general, which is expected to be applicable 
to other situations and other SG systems.

Before closing this subsection, we mention the physical consequence of the absence of the zeros in the SG phase in the present situation. From the point of view of the chaos effect, this is quite natural since the present model does not show the temperature chaos~\cite{Rizzo:06}. Generally, the chaos effect is connected to the change of the dominating pure states when we vary the corresponding physical parameter. For the spherical SG model, it is shown that the pure state essentially does not change in temperature below the SG transition point. This means the absence of the temperature chaos, and of a certain transitions mentioned in section~\ref{sec:introduction}. To provide a clear connection between the chaos effect and the DOZ, we need to investigate models exhibiting the chaos effect. One of the simplest choice of such models is the spherical SG model with multiple many-body interactions~\cite{Rizzo:06}, which is examined in the next subsection. 

\subsection{Multiple interaction case}

The Hamiltonian of the $(p+r)$-body interacting spherical model is given by
\be
\mathcal{H}=-\sum_{i_1<\cdots<i_p}J_{i_1 \cdots i_p}S_{i_1}\cdots S_{i_p}
-\epsilon\sum_{i_1<\cdots<i_r}K_{i_1 \cdots i_r}S_{i_1}\cdots S_{i_r}.
\ee
As \eref{Jij}, the coupling constants $J_{i_1 \cdots i_p}$ and 
$K_{i_1 \cdots i_r}$ are Gaussian variables with 
the variances $J^2p!/2N^{p-1}$ and $J^2r!/2N^{r-1}$, respectively. 

The calculation of $\phi(\beta,n)$ of this system is the same as 
given in section~\ref{sec:Replica}. 
The resultant expression of $\phi(\beta,n)$ is obtained by 
just replacing $q_{ab}^{p}$ with $(q_{ab}^{p}+\epsilon^2 q_{ab}^{r})$ 
in \eref{phi-sp}, and $(q'_{ab})^p$ and $\tilde{q}_{ab}^p$ are replaced in a similar manner. 
The analysis to obtain the phase diagram and the DOZ is also 
the same and we omit it here. 
We just give the result below. 

\begin{figure}[htbp]
\begin{center}
\includegraphics[width=0.5\columnwidth]{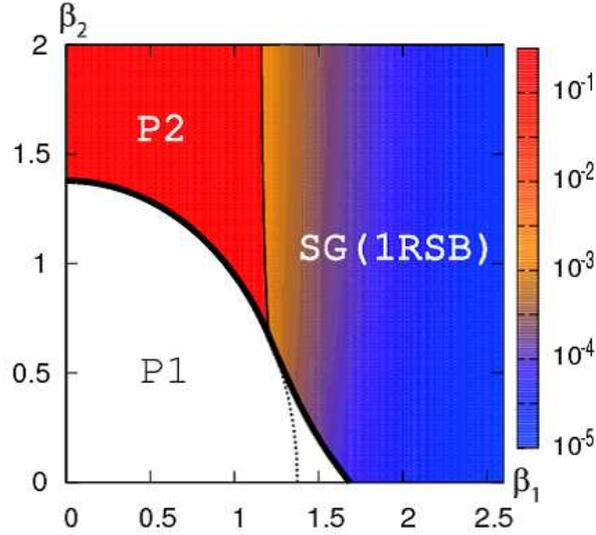}
\caption{Complex-temperature phase diagrams of the $(p+r)$-body spherical SG for $p=3$ and $r=4$ with $J=1$ and $\epsilon=0.2$. The DOZ takes finite values on the P2 and SG phases and is coloured in a logarithmic scale. The one-dimensional DOZ becomes finite on bold phase boundaries as the single $p=3$ case.} 
\label{fig:PD-multiple}
\end{center}
\end{figure}

As an example, we show the case $(p,r)=(3,4)$ and $\epsilon=0.2$, where the 1RSB solution is correct~\cite{Rizzo:06}. The shape of the phase diagram is almost the same as that of the usual $p=3$ spherical SG model, and the corresponding critical temperatures are $\beta_c=1.68462/J$ and $\beta_p=1.37877/J$. A remarkable property in this case is that the DOZ in the SG phase is finite. In figure \ref{fig:PD-multiple}, we give the DOZ coloured in a logarithmic scale on the phase diagram. 
We can find that the DOZ values of the SG phase are rather small in comparison with the P2 one: The typical values are $\rho=O(10^{-1})$ and $\rho \leq O(10^{-3})$ in the P2 and SG phases, respectively. This means that the DOZ discontinuously changes at the P2-SG boundary as for the single $p$ case. 

\begin{figure}[htbp]
\begin{center}
\includegraphics[height=0.3\columnwidth,width=0.46\columnwidth]{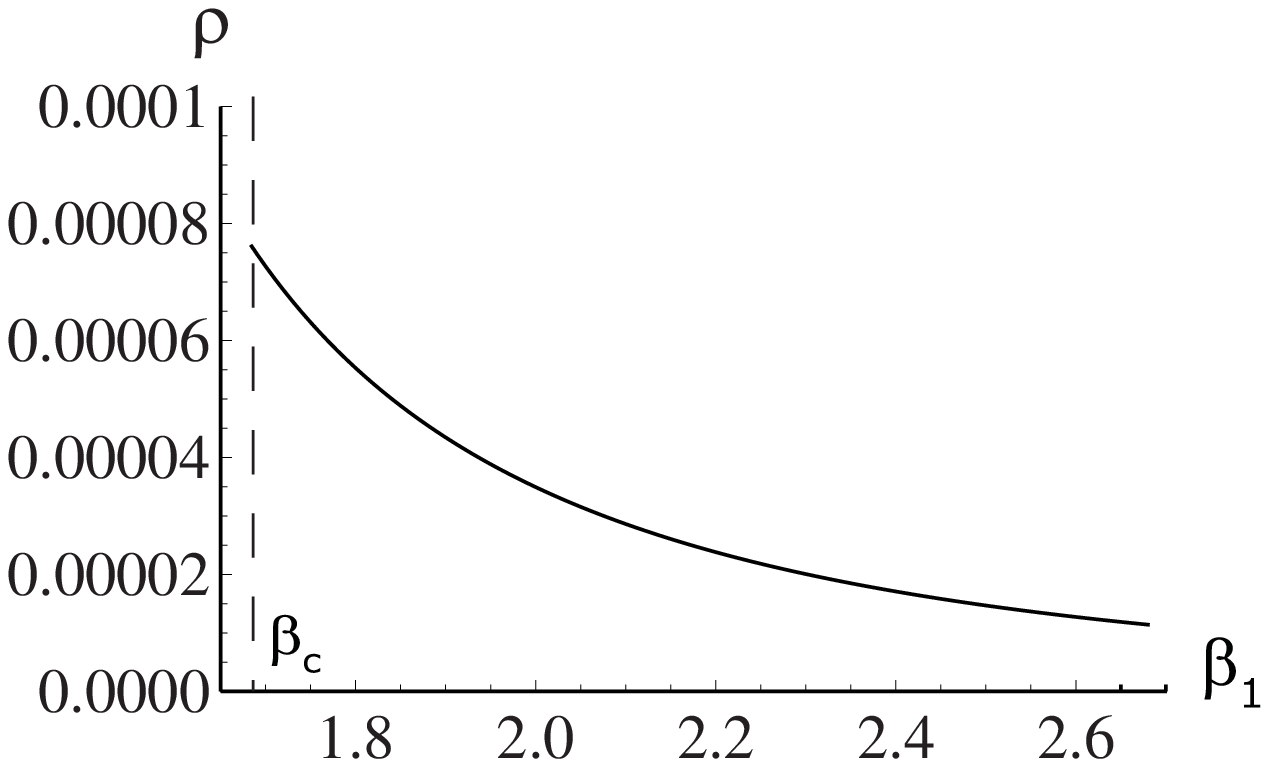}
\hspace{2mm}
\includegraphics[height=0.3\columnwidth,width=0.46\columnwidth]{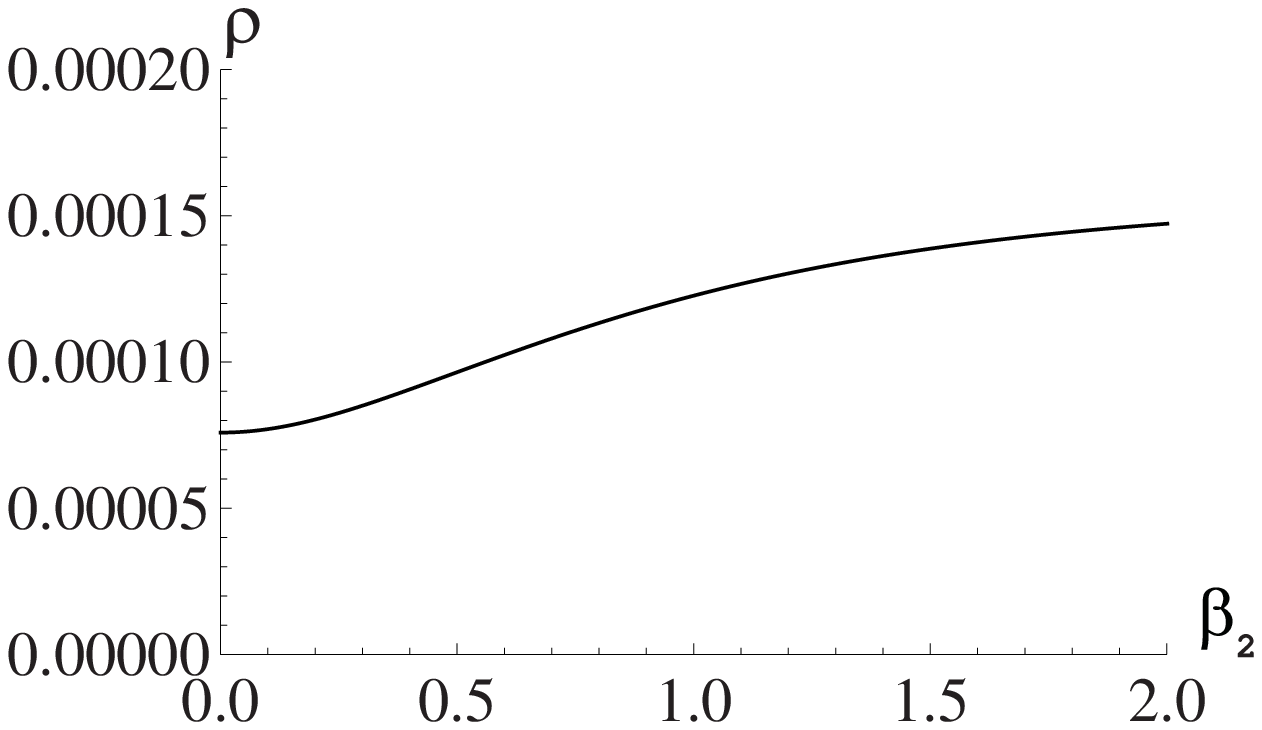}
 \caption{The DOZ on the real axis (left) and on a vertical line at $\beta_1=\beta_c$ (right) in the SG phase of the case $(p,r)=(3,4)$ and $\epsilon=0.2$. 
}
 \label{fig:rho-SG}
\end{center}
\end{figure}

Obviously, the DOZ on the real axis is more important, and we plot it in the left panel of figure \ref{fig:rho-SG}. The figure clearly shows that the DOZ on the real axis is finite. This is quite contrast to the case of the single $p$-body interaction, which strongly suggests that the DOZ on and around the real axis in the SG phase is closely related to the temperature chaos. We further discuss about this point in the next subsection. For quantitative comparison, we also give the DOZ on a vertical line at the critical point $\beta=\beta_c$ in the right panel of figure \ref{fig:rho-SG}. 
From these figures, we can see that the density tends to decrease/increase as the real/imaginary part of $\beta$ grows. Although it is interesting to see the limiting value of $\rho$ in the limit $\beta_2 \to \infty$ in the right panel of figure~\ref{fig:rho-SG}, we could not find appropriate solutions for large $\beta_2$ due to some uncontrollable technical difficulties.

\subsection{Discussion}

So far, we have investigated the zeros of the spherical SGs in several situations. 
For the $p=2$ case where the RS solution is correct in all the region, 
the zeros distribute two-dimensionally in the P2 phase, 
though the SG phase has no zeros. 
The zeros are also absent in the 1RSB-SG phase for the $p=3$ case. 
These observations indicate that the zeros are not directly related to the RSB. 
On the other hand, in the SG phase of the $(p+r)$-body interacting 
spherical model, the DOZ takes nonzero values. 
Since this system shows the temperature chaos being absent 
in the single $p$-body interaction case, 
we can naturally speculate that the chaos effect is closely related 
to the DOZ distributed in the SG phase, 
especially the values of DOZ on the real axis is quite important.  

The above statement becomes clearer by referring the DOZ of 
the REMs~\cite{Takahashi:11}. 
For the generalized REM (GREM) in the continuum limit of the hierarchy,
the one-dimensional DOZ on the phase boundaries are accumulated 
to become a two-dimensional distribution. 
That two-dimensional DOZ looks similar to the $(p+r)$-body 
spherical case, but a crucial difference is that the DOZ of the GREM
is zero on the real axis of $\beta$.
This can be interpreted as follows. 
In the GREM, by definition, the consecutive transitions in temperature 
are characterized by a sequential freezing of a part of 
the total spins~\cite{Obuchi:10}. On one hand,
this means that the nature of the transitions essentially becomes 
of second order, which explains the absence of zeros directly 
on the real axis. 
On the other hand, the freezing character of the transitions implies 
the presence of correlations among the equilibrium spin configurations 
at different temperatures, which leads to the absence of 
the temperature chaos as shown in~\cite{Franz:95}. 
Combining this observation and the fact that the chaos effect is connected to discontinuous changes of the dominant part of pure states~\cite{Rizzo:06}, we can reasonably conclude that the chaos effect generally has a first-order-transition-like nature and is signaled by the zeros on the real axis.

We here refer to the Sherrington-Kirkpatrick (SK) model~\cite{Sherrington:75} known to show the temperature chaos~\cite{Rizzo:03}. In this model, although it is known that there emerge many pure states in the SG phase, properties of the pure states are quite unclear in comparison with those of the spherical models. This may be regarded as a consequence of a local instability of the phase space: The so-called de Almeida-Thouless instability~\cite{Almeida:78}. This instability directly causes the SG transition of the SK model, and accordingly the nature of the transition becomes continuous, which may question the above description of the chaos effect. However, considering Bethe SGs expected to show the essentially same behavior as the SK model, we can find that the zeros distribute in the SG phase even on the real axis of temperature~\cite{Matsuda:10}. This fact implies that the discontinuous change of the dominant pure states also occurs in the SG phase of the SK model, which supports our above description.

We should also notice that the zeros on the real axis in the SG phase do not mean the singularities of the free energy. Actually, the free energy of this model for real $\beta$ does not show any singularities in the SG phase. This is because the DOZ are two-dimensionally distributing in the SG phase. An interpretation of this fact can be naturally obtained by considering the analogy with the electrostatics explained in section~\ref{sec:zeros}. We can see that the electrostatic potential can be analytic in a region where the point charges distribute two-dimensionally, which also means the absence of the singularities of the free energy in a region where the zeros distribute two-dimensionally.

Conversely, partition-function zeros can detect extraordinary behaviour even not appearing as singularities of the free energy. We stress that our result in this paper becomes the first evidence and demonstration of that fact. 

\section{Conclusion}

In this paper, we investigated the zeros in the complex temperature 
plane of the many-body interacting spherical SGs, in the single $p$-body and multiple $(p+r)$-body interacting cases. Our formulation utilizes the replica method and generalizes the Parisi scheme to be applicable in the complex-parameter case. The relations between the pure-state structure and the overlap matrices were also considered. Based on the formulation, we derived the phase diagrams in the complex-temperature plane and calculated the DOZ on each phase and boundary. By changing the parameters $p$ and $r$, we could easily investigate several physically-different situations and examined the possible relations among the DOZ, the RSB and the chaos effect. The notable significances of the result are as follows:
\begin{itemize}
\item{The RSSG phase cannot have the finite DOZ.}
\item{The RSB is not necessarily connected to 
the two-dimensionally distributed zeros.}
\item{The two-dimensionally distributed zeros around and 
on the real axis are closely related to the chaos effect, 
and do not necessarily lead to singularities of the free energy.}
\end{itemize}

The formulation presented in this paper has some possible applications. One of the most simple applications is to investigate the SG systems with the FRSB such as the SK model. Although we constructed our solution in the 1RSB level, it is possible to extend the solution to the FRSB. The DOZ in the FRSB phase of the Bethe SG was studied in~\cite{Matsuda:10}, but the RS ansatz was used to calculate the DOZ, which can involve potential errors in the estimation. Besides, in that result we cannot see the discrimination between the P2 and FRSB-SG phases, which is clearly different from our present result. This difference may be due to the RS ansatz~\cite{Matsuda:10}, or due to the difference between the 1RSB and FRSB. It is also a question how the de Almeida-Thouless instability~\cite{Almeida:78} relates to the DOZ. These questions motivate us to investigate the FRSB-SG phase by our current formulation, which will be our future work. 

In the presented paper, we extracted some physical properties of the zeros only on and around the real axis, but the DOZ in the whole complex plane potentially has more significance. The P2 phase has a particular interest since the DOZ on and around the imaginary axis is possibly related to some dynamical properties~\cite{Bena:05,During:10}. Hence, it will be interesting to seek physical contents of the DOZ in the whole complex plane. Other possible applications, such as to finite-dimensional SGs~\cite{Bray:87} and structural glasses~\cite{Mezard:99}, are also important directions. We hope that the presented formulation and result inspire those researches, and lead to revealing origins of extraordinary behaviour of glassy systems.

\section*{Acknowledgments}

The authors are grateful to Y Matsuda and H Yoshino for useful discussions. TO is supported by a Grant-in-Aid Scientific Research on Priority Areas `Novel State of Matter Induced by Frustration' (19052006 and 19052008). A part of numerical computations in this work were carried out at the Yukawa Institute Computer Facility.

\appendix
\section{Solving the equations of state with complex parameters}
\label{sec:app1}
Here, we describe how to solve the complex equations of state 
and to obtain the phase boundaries in figure~\ref{fig:PD}. 

We start from the simpler case, i.e. the boundary 
between the P2 and 1RSB-SG phases. 
As explained in section~\ref{sec:PDp=3}, the condition $m=1$ is 
essential to determine the P2-1RSB boundary. 
Under this condition, the generating functions $g_{\rm P2}$ 
and $g_{\rm 1RSB}$ automatically becomes identical. 
This means that $g_{\rm 1RSB}$ becomes independent from $q_1$ and $q'_1$, 
and the same is true for the equation of state 
of $\tilde{q}_1$, \eref{EOS-1RSB-qtil1}. 
On the other hand, the variational condition 
with respect to $m$, \eref{EOS-1RSB-m}, which should be satisfied 
at the phase boundary, still depends on $q_1$ and $q'_1$. 
Hence, to obtain the phase boundary we need to calculate $q_{1}$ 
through \eref{EOS-1RSB-q1}, though it is not needed to 
evaluate $g_{\rm 1RSB}$ at $m=1$. 
Summarizing these observations, we adopt the following procedures 
to obtain the boundary between the P2 and 1RSB-SG phases:
\begin{enumerate}
\item{Fix a value of $\mu=p\beta^2J^2/2$.}
\item{Calculate $\tilde{q}_1$ through \eref{EOS-1RSB-qtil1} with $m=1$. 
This can be easily performed by usual methods such as iteration, 
bisection method and Newton's method. 
For $p=3$, even the analytic solution can be obtained.}
\item{Calculate $q_{1}$ by solving \eref{EOS-1RSB-q1} with substitution 
of the above $\tilde{q}_1$ under $m=1$ (note that $q'_1=q_1^*$ ). 
Since $q_1$ is complex, the iteration and bisection methods do not work well. 
We employ Newton's method with an appropriately-chosen 
initial value of $q_{1}$. 
Empirically, we find that the initial value should have 
a small imaginary part and a real part slightly smaller than unity.}
\item{Evaluate the left-hand side of \eref{EOS-1RSB-m} with $m=1$ 
by using the obtained values of $\tilde{q}_1$ and $q_{1}$. 
If the value is sufficiently small, the given $\beta=\sqrt{2\mu/p}/J$ 
gives the desired phase boundary. 
Otherwise, restart from (i) with a new value of $\mu$. 
To efficiently search the boundary, we actually fix $\Im{\beta}$ 
and gradually change $\Re{\beta}$ to find a value of $\beta$ 
at which \eref{EOS-1RSB-m} is satisfied.
}
\end{enumerate}
These procedures actually work well and the P2-1RSB boundary 
is obtained straightforwardly. 

Next, we consider the boundary between the P1 and 1RSB-SG phases. 
For this, we need to treat all the equations of state 
\eref{EOS-1RSB-q1}-\eref{EOS-1RSB-m} with mutual dependence among 
the order parameters, unlike the P2-1RSB case. 
To actually solve \eref{EOS-1RSB-q1}-\eref{EOS-1RSB-m}, 
we focus on the fact that $\tilde{q}_1$ and the left-hand side 
of \eref{EOS-1RSB-m} are real. 
This property enables us to use the bisection method 
to evaluate those two equations. 
The resultant procedures we accept are as follows:
\begin{enumerate}
\item{Fix a value of $\mu=p\beta^2J^2/2$.}
\item{Calculate $q_1$, $\tilde{q}_1$ and $m$. 
Call the bisection subroutine with respect to $m$ by \eref{EOS-1RSB-m}.
\begin{enumerate}
\item{Fix three appropriate values of $m$, $m^h>m^l$  and $m^m=(m^h+m^l)/2$.}
\item{Calculate $q_1$ and $\tilde{q}_1$ for given three values of $m$. 
For this, call the bisection subroutine with respect to $\tilde{q}_1$ 
by \eref{EOS-1RSB-qtil1}.
\begin{enumerate}
\item{Fix three appropriate values of $\tilde{q}_1$, 
$\tilde{q}_1^h>\tilde{q}_1^l$  
and $\tilde{q}_1^m=(\tilde{q}_1^h+\tilde{q}_1^l)/2$.}
\item{Calculate $q_1$ for given $m$ and three values of $\tilde{q}_1$ 
by solving \eref{EOS-1RSB-q1} and obtain $(q_1^h,q_1^m,q_1^l)$. 
Newton's method with an appropriate initial value of $q_{1}$ is again useful.}
\item{Compare the left-hand side of \eref{EOS-1RSB-qtil1} 
with substitutions of $(q_1^h,\tilde{q}_1^h)$, $(q_1^m,\tilde{q}_1^m)$ 
and $(q_1^l,\tilde{q}_1^l)$. 
Replace $\tilde{q}_1^h$ or $\tilde{q}_1^l$ with $\tilde{q}_1^m$ 
depending on the compared values of \eref{EOS-1RSB-qtil1}}.
\item{Repeat 1-3 until the value of $\tilde{q}$ converges. 
Return the resultant $(q_1,\tilde{q}_1)$.}
\end{enumerate}
}
\item{Compare the left hand side of \eref{EOS-1RSB-qtil1} 
with substitutions of $(m^h,q_1^h,\tilde{q}_1^h)$, 
$(m^m,q_1^m,\tilde{q}_1^m)$ and $(m^l,q_1^l,\tilde{q}_1^l)$. 
Replace $m^h$ or $m^l$ with $m^m$ depending on the compared values 
of \eref{EOS-1RSB-m}.
}
\item{Repeat (a)-(c) until the value of $m$ converges. 
Return the resultant $(m,q_1,\tilde{q}_1)$.}
\end{enumerate}
}
\item{Compare the values of $g_{\rm P1}$ and $g_{\rm 1RSB}$ 
by using the obtained $(m,q_1,\tilde{q}_1)$. 
If the difference is sufficiently small, 
the given $\mu$ gives the phase boundary. 
Otherwise, restart from (i) with a new value of $\mu$.}
\end{enumerate}
To actually conduct these procedures, some difficulties are involved 
in choosing the appropriate values of $\tilde{q}_1$ and $m$ 
for the bisection subroutines. 
If the chosen values are inappropriate, the converged values become unphysical. 
To resolve this point, we start from the critical point $\beta=\beta_c$ 
and gradually change the value of $\beta$ step by step, 
with using the values of order parameters in the previous step 
as the initial values of $\tilde{q}_1$ and $m$ in the current step. 
Although this prescription works well for assessing the phase boundaries, 
the trouble becomes more serious when we evaluate the DOZ 
in a region far from the real axis in the SG phase, due to 
bad behaviour of \eref{EOS-1RSB-q1}-\eref{EOS-1RSB-m} in that region. 
A more effective routine to solve the complex equations 
of state \eref{EOS-1RSB-q1}-\eref{EOS-1RSB-m} will quite benefit 
to assess the DOZ, but in the presented results we did not pursue this point and 
just tuned the parameters until physically-plausible results are obtained.


\end{document}